# Pressure-Induced Superconductivity in Iron-Based Spin-Ladder Compound BaFe$_{2+\delta}$(S$_{1-x}$Se$_x$)$_3$

**Hiroki Takahashi** [1,*], **Ryosuke Kikuchi** [1], **Chizuru Kawashima** [1], **Satoshi Imaizumi** [2], **Takuya Aoyama** [2] **and Kenya Ohgushi** [2]

College of Humanities & Sciences, Nihon University, Tokyo 156-8550, Japan; rkikuchi@vibra.co.jp (R.K.); kawashima.chiduru@nihon-u.ac.jp (C.K.);
Graduate School of Science, Tohoku University, Sendai 980-8578, Japan; satoshi.imaizumi.r7@dc.tohoku.ac.jp (S.I.); aoyama@tohoku.ac.jp (T.A.);
kenya.ohgushi.d4@tohoku.ac.jp (K.O.)
Correspondence: takahashi.hiroki@nihon-u.ac.jp

**Abstract:** The iron-based superconductors had a significant impact on condensed matter physics. They have a common structural motif of a two-dimensional square iron lattice and exhibit fruitful physical properties as a strongly correlated electron system. During the extensive investigations, quasi-one-dimensional iron-based spin-ladder compounds attracted much attention as a platform for studying the interplay between magnetic and orbital ordering. In these compounds, BaFe$_2$S$_3$ and BaFe$_2$Se$_3$ were found to exhibit superconductivity under high pressure, having a different crystal and magnetic structure at low temperature. We report a brief review of the iron-based spin-ladder compound and recent studies for BaFe$_{2+\delta}$(S$_{1-x}$Se$_x$)$_3$. BaFe$_2$(S$_{0.75}$Se$_{0.25}$)$_3$ is in the vicinity of the boundary of two different magnetic phases and it is intriguing to perform high pressure experiments for studying superconductivity, since effects of large magnetic fluctuations on superconductivity are expected. The effect of iron stoichiometry on the interplay between magnetism and superconductivity is also studied by changing the iron concentration in BaFe$_{2+\delta}$Se$_3$.

## 1. Introduction

### 1.1. Iron-Based Spin-Ladder Material

The iron-based superconductor had a significant impact on condensed matter physics as a strongly correlated electron system based on the two-dimensional square iron lattice and exhibits characteristic magnetic phases next to the superconducting phase. These compounds triggered the extensive studies in the interplay between magnetism, orbital ordering, and superconductivity as a new material platform for further exploration of high-$T_C$ superconductors [1–4]. A stripe-type magnetic ordering is observed in the 1111, 122, 111, and 11 type iron-based superconductors [1,2,5–7], and a block-type magnetic ordering is observed in 245 type iron-based superconductor [8,9]. During the extensive investigations for these compounds an iron-based spin-ladder compound attracted much attention due to having a different dimensional property. The iron-based spin-ladder compounds AFe$_2$X$_3$ (A = Ba, K, Cs; X = S, Se, Te) have a quasi-one-dimensional two-leg ladder formed by edge-sharing [FeX$_4$] tetrahedral structure with channels occupied by A cations, as shown in Figure 1a [10]. These compounds exhibit characteristic crystal structure and magnetic ordering [11–20]. The magnetic structures of BaFe$_2$S$_3$ and AFe$_2$Se$_3$ (A = K,

Cs) are stripe-type, in which the magnetic moments couple ferromagnetically along the rung direction, and antiferromagnetically along the leg direction [16–19]; the magnetic structure of BaFe$_2$Se$_3$ is block-type, in which the magnetic moments form ferromagnetic Fe$_4$ units and couple antiferromagnetically along the leg direction [15–19]. These magnetic structures shown in Figure 1b are one-dimensional analogue of stripe and block type magnetism observed in the two-dimensional iron-based superconductors. Since there are some similarities between the two-dimensional iron-based system and one-dimensional spin-ladder system, much attention has been paid for experimental and theoretical study of the iron-based spin-ladder compounds. Several years ago, pressure-induced superconductivity was found in the iron-based spin-ladder compounds BaFe$_2$S$_3$ and BaFe$_2$Se$_3$ [21,22].

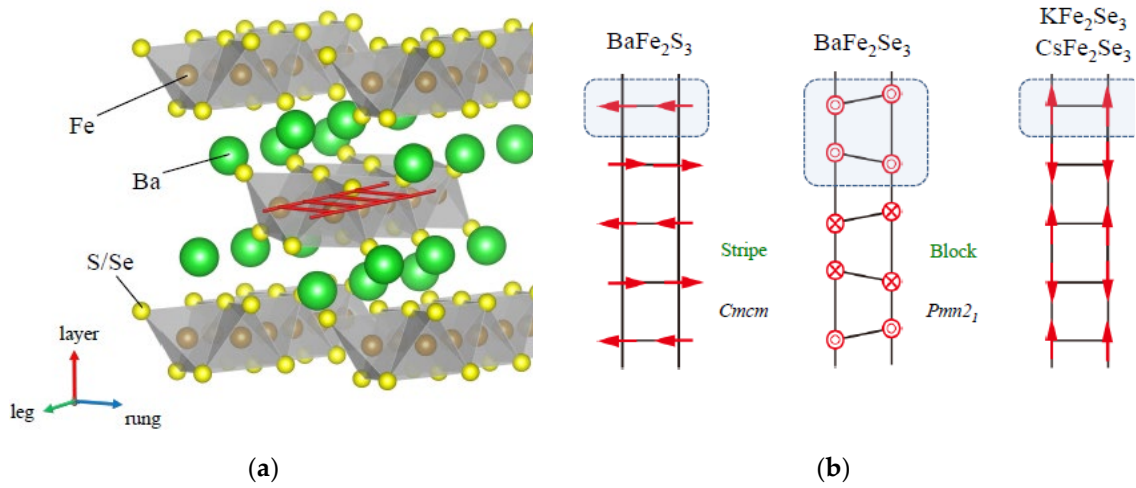

(a) (b)

**Figure 1.** (**a**) Crystal structure of iron-based spin-ladder compounds BaFe$_2$S$_3$ and BaFe$_2$Se$_3$, consisting of edge-shared Fe(S/Se)$_4$ tetrahedra extending along the "leg direction" and of the channel occupied by Ba atoms, in which Fe atoms form the ladder structure shown in this figure [10]. (**b**) Magnetic structures in the ladder of BaFe$_2$S$_3$, BaFe$_2$Se$_3$, and AFe$_2$Se$_3$ (A = K, Cs). In stripe-type magnetic ordering, the magnetic moments couple ferromagnetically along the rung direction and antiferromagnetically along the leg direction, and in block-type magnetic ordering, the magnetic moments form ferromagnetic Fe$_4$ units and couple antiferromagnetically along the leg direction.

*1.2. Spin-Ladder Compound BaFe$_2$S$_3$*

Figure 2 shows a $P$-$T$ phase diagram of BaFe$_2$S$_3$ [21,23,24]. BaFe$_2$S$_3$ shows the insulating behavior at ambient pressure caused by the electron correlation effect, though metallic behavior is expected in the unfilled 3d manifold. High-pressure neutron diffraction measurements revealed that the $T_N$ increases steeply from 100 K to 160 K with applying pressure and decreases gradually with further applying pressure [24]. It was indicated that some sort of quantum phase transition occurs around 1 GPa [24]. Electrical resistivity measurements also revealed the pressure dependence of $T_N$, in which an increase of $T_N$ is also observed below 3 GPa [23]. In electrical resistivity measurements, a weak anomaly is observed at $T^* \sim 180$ K, which is considered to be related to the orbital ordering, because nematic susceptibility measurements revealed a distinct change at $T^*$ [25]. Since the characteristic temperature $T^*$ is merged into $T_N$ at ~1 GPa in the high-pressure resistivity measurements [23], the quantum phase transition observed at 1 GPa seems to be related to the interplay between the magnetic and orbital ordering. With applying pressure using a diamond anvil cell (DAC), the insulating properties are gradually suppressed, and the electrical resistance exhibits a metal-insulator transition at 11 GPa. Simultaneously, the resistance drops steeply at 14 K, which is originating from superconductivity [21]. The superconducting transition temperatures ($T_c$) depicted from several samples, whose $T_N$ values are ranging from 105 to 124 K, show a dome-shape phase boundary in the pressure range from 10 to 16 GPa. The absence of zero resistance could be due to technical limitations inherent in DAC, as fully hydrostatic compressive stress could not be applied in the sample space of DAC when using a solid pressure-transmitting medium. Using a cubic anvil press (CAP), zero resistance was observed and $T_c$ was reported to reach 24 K [23]. The shielding volume fraction of BaFe$_2$S$_3$ is confirmed to be about 64% at 5 K by AC susceptibility measurement, which indicates that the superconductivity has a bulk origin.

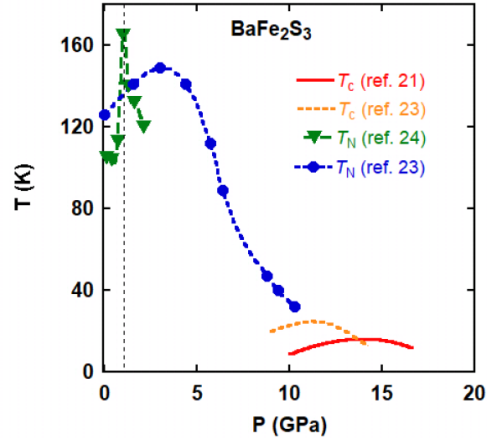

**Figure 2.** Pressure-temperature electronic phase diagram for BaFe$_2$S$_3$. The superconducting transition temperature ($T_c$) is determined by electrical resistivity measurement using a diamond anvil cell [21] and using a cubic anvil press [23]. The antiferromagnetic transition temperature ($T_N$) is determined by neutron diffraction measurements [24] and electrical resistivity measurement [23]. The dotted line shows the pressure where the $T_N$ and magnetic ordered moment exhibit abrupt increases [24].

The crystal structure under high pressure of BaFe$_2$S$_3$ was obtained by X-ray diffraction measurements [21,26]. The linear compressibility is reported to be $\kappa_a$ = 4.6 × 10$^{-3}$ (rung direction), $\kappa_b$ = 6.5 × 10$^{-3}$ (interlayer direction), and $\kappa_c$ = 4.0 × 10$^{-3}$ (leg direction). No structural phase transition has been observed, indicating that the superconductivity appears in the ladder structure. The $b$ axis perpendicular to the ladder layer is most compressible, and it is likely that the metal-insulator transition and superconductivity are caused by increasing charge transfer of the Fe 3$d$ electrons across the inter-ladder units.

### 1.3. Spin-Ladder Compound BaFe$_2$Se$_3$

BaFe$_2$Se$_3$ shows the crystal structure of non-centrosymmetric orthorhombic $Pmn2_1$ symmetry below 400 K, which means this phase loses spatial inversion symmetry. This phase exhibits a multiferroic property, which was confirmed by second-harmonic generation experiments [27]. With increasing temperature, this phase transforms to orthorhombic $Pnma$ above 400 K ($T_{s2}$) and to orthorhombic $Cmcm$ at 660 K ($T_{s1}$) [28]. The structural transition from orthorhombic $Pnma$ to orthorhombic $Cmcm$ is also observed under a high pressure of about 4 GPa at ambient temperature [29]; it is reported that the tilting angle of iron ladders disappears in this transition. Pressure-induced superconductivity was observed under pressure of 11.5 GPa and the maximal $T_c$ is about 11 K [22]. Although the pressure dependence of magnetic structure and $T_N$ are not experimentally known, Fe $K\beta$ X-ray emission spectroscopy exhibits that the magnetic moment is continuously suppressed and superconductivity appears when the magnetic moment is sufficiently suppressed [22]. The theoretical calculation indicates that the magnetic structure is replaced from block-type to stripe-type at ~12 GPa and magnetic moments vanish at 30 GPa [30]. Thus, the superconductivity of BaFe$_2$Se$_3$ appears under the existence of magnetic moment, and it is likely that the mechanism of superconductivity is different from BaFe$_2$S$_3$. Coupled with theoretical calculations, it is interesting to carry out a high-pressure experiment to confirm whether the additional superconducting phase appears or not above 30 GPa at which the magnetic moment disappears.

1.4. Spin-Ladder Compound BaFe$_2$(S$_{1-x}$Se$_x$)$_3$

In the previous section, we indicate that BaFe$_2$S$_3$ and BaFe$_2$Se$_3$ exhibit pressure-induced superconductivity with a different electronic and magnetic state since these compounds show a different crystal and magnetic structure. BaFe$_2$(S$_{1-x}$Se$_x$)$_3$ is regarded as a bandwidth-control system [31]. Electrical resistivity measurements exhibit that BaFe$_2$S$_3$ is more metallic than BaFe$_2$Se$_3$, since the replacement of S$^{2-}$ ions by Se$^{2-}$ ions induces negative chemical pressure on the system, considering only the ionic radius. Figure 3 shows an $x$-$T$ electronic phase diagram for BaFe$_2$(S$_{1-x}$Se$_x$)$_3$ [31]. The magnetic structure transforms from the stripe-type to the block-type at $x$ = 0.23 without any intermediate phase and the $T_N$ value is suppressed at the transition composition due to the critical nature between two different kinds of magnetic orderings. The transition temperature $T^*$ and $T_{s2}$ values related to orbital ordering show a similar $x$ dependence on $T_N$, which indicates a close relation between the magnetic and the orbital ordering. We focus on the magnetic phase boundary at $x$ = 0.23 to study the superconductivity under high pressure. It is interesting to examine what kind of superconductivity appears under high pressure.

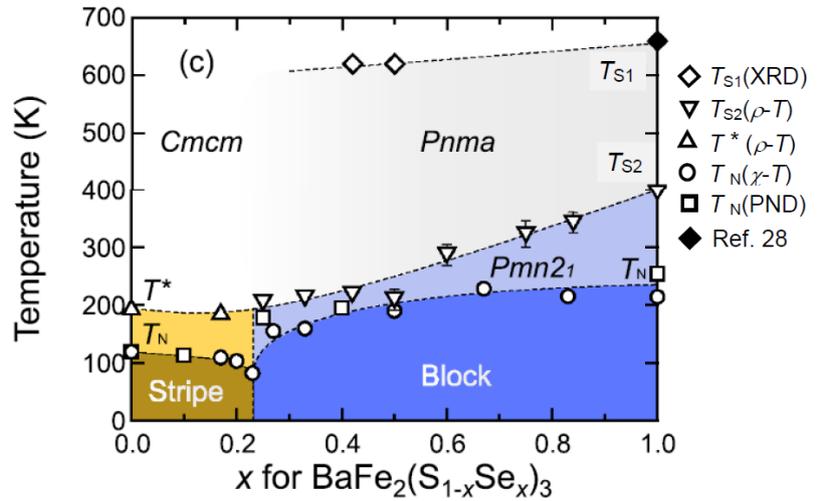

**Figure 3.** Electronic phase diagram of BaFe$_2$(S$_{1-x}$Se$_x$)$_3$ taken from Ref. [31]. $T_{s1}$ is the structural transition temperature from *Cmcm* to *Pnma*. The composition $x$ = 0.23 is the boundary of two different magnetic phases. $T_{s2}$ and $T^*$ are orbital ordering temperature. The orbital ordering and magnetic ordering temperature are suppressed at this composition.

Concerning the solid solution, the filling-control system Ba$_{1-x}$Cs$_x$Fe$_2$Se$_3$ is studied for the interplay between magnetism and superconductivity [32]. Superconductivity has not been found in this system under high pressure, though no magnetic ordering phase appears between the block-type BaFe$_2$Se$_3$ and stripe-type CsFe$_2$Se$_3$ magnetic phase.

*1.5. Purpose of This Study*

We study a BaFe$_2$(S$_{1-x}$Se$_x$)$_3$ system to elucidate the interplay between magnetic and orbital ordering, and superconductivity under high pressure, and focus on the $x$ ~ 0.23 composition close to the magnetic phase boundary. At the magnetic phase boundary where the $T_N$ is already decreased at the ambient pressure, the appearance of superconductivity is expected because novel superconductivity is often observed at a quantum critical point in the strongly correlated electron systems. In BaFe$_2$S$_3$, $T_N$ of stripe-type magnetic ordering shows a significant dependence on the synthesis condition, in which the $T_N$ values of 119–124 K was obtained. Moreover, $T_c$ changes with similar trend to the value of $T_N$. We therefore study the $\delta$ dependence of $T_c$ and $T_N$ in BaFe$_{2+\delta}$Se$_3$ to examine the effect of the stoichiometry on the pressure-induced superconductivity. We examine the existence or absence of pressure-induced superconductivity in BaFe$_{2+\delta}$Se$_3$ above 30 GPa, based on the theoretical calculation.

## 2. Materials and Methods

Single crystals of BaFe$_2$(S$_{1-x}$Se$_x$)$_3$ ($x$ = 0.25) and BaFe$_{2+\delta}$Se$_3$ ($\delta$ = 0, 0.1, and 0.2) were synthesized by the slow cooling method [33]. Starting materials Ba, Fe, S, and Se were mixed according to the stoichiometric ratio, and the mixture was put into a carbon crucible. The crucible was sealed in a quartz tube. The quartz tube was heated at 1150 °C for 40 h and slowly cooled down to 750 °C for typically 60 h. The values of $x$ and $\delta$ are nominal compositions. In BaFe$_{2+\delta}$S$_3$, the iron composition evaluated by SEM image is smaller than the ideal value expected from the chemical formula [34]. To compensate for this defect and obtain a rather stoichiometric sample, an excess of iron is added to the starting materials in BaFe$_{2+\delta}$S$_3$. We presume that the same tendency is expected for BaFe$_2$Se$_3$. However, it is indicated that the Fe-rich inclusion appears in the Fe excess BaFe$_{2+\delta}$S$_3$ [35]. It is observed that the slight increase of ferromagnetic signal was detected with increasing excess Fe in our SQUID measurements in BaFe$_{2+\delta}$Se$_3$. Figures 4 shows the powder X-ray diffraction profiles of BaFe$_{2+\delta}$Se$_3$ ($\delta$ = 0, 0.1, and 0.2). The measurements were carried out at room temperature using a RIGAKU Smartlab diffractometer (RIGAKU, Akishima, Japan) with Cu K$\alpha$ source. One can find that almost all peaks can be attributed to the main phase, although there is a tiny impurity peak around 32 degrees in the profile of $\delta$ = 0.2, which seems to be due to ferromagnetic Fe$_7$Se$_8$. The inset of Figure 4 is the optical image of BaFe$_2$Se$_3$. Typical dimension of obtained crystal is 1 mm × 3 mm × 0.2 mm, where the longest direction is along the ladder.

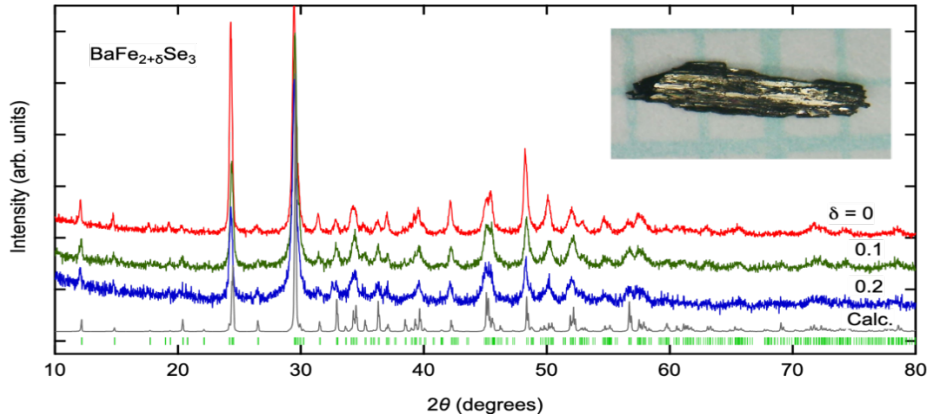

**Figure 4.** Powder X-ray diffraction patterns for BaFe$_{2+\delta}$Se$_3$. The gray curve and green ticks represent calculated intensities and expected peak positions of BaFe$_2$Se$_3$, respectively. Inset is the optical image of a BaFe$_2$Se$_3$ crystal on graph paper ruled into 1-mm squares.

The magnetic susceptibility and electrical resistivity were measured using a commercial setup (Quanrum Design, San Diego, USA) (MPMS and PPMS, Quantum Design). The $T_N$s of BaFe$_{2+\delta}$Se$_3$ in this study are 242 K, 241 K, and 245 K for $\delta$ = 0, $\delta$ = 0.1, and $\delta$ = 0.2, respectively. Electrical resistivity measurements were carried out by a standard dc four-probe method. A DAC made of CuBe alloy was used for electrical resistance measurements at pressures up to 40 GPa.

The sample chamber was filled with powdered NaCl as the pressure-transmitting medium, using a rhenium gasket. Thin (10 μm thick) platinum ribbons were used as leads for the standard dc four-probe analysis. Each rectangular sample was 0.1 × 0.1 mm and 0.03 mm thick. A thin BN layer acted as electric insulation between the leads and the rhenium gasket and finely ground ruby powder scattered in the sample chamber was used to determine the pressure by the standard ruby fluorescence method.

## 3. Results and Discussion

### 3.1. Electrical Resistivity under High Pressure for Spin-Ladder Compound $BaFe_2(S_{1-x}Se_x)_3$ (x = 0.25)

As shown in Figure 3, $BaFe_2(S_{0.75}Se_{0.25})_3$ is in the vicinity of the phase boundary between stripe-type and block-type magnetic ordering. Figure 5a shows the temperature dependence of the electrical resistance of $BaFe_2(S_{0.75}Se_{0.25})_3$ along the leg direction. The metal-insulator transition is observed around 7–10 GPa, which is similar to the one of $BaFe_2Se_3$. Figure 5b shows the $R(T)$ curve of metallic state above 10 GPa and Figure 5c shows the low-temperature results. A sudden decrease of resistance is observed around 10 K which is a rather small change compared with the one of $BaFe_2Se_3$. This decrease of resistance is thought to be attributed to the superconducting transition. As shown in Figure 5d, the superconducting transition at 12 GPa shifts to the low-temperature side with increasing electric current, which is consistent with superconducting characteristics. Since a zero resistance could not be observed in the $R(T)$ curve, the superconducting temperature of this compound is expressed by $T_{co}$ for the onset of the apparent superconducting transition. Figure 5e shows the pressure dependence of $T_{co}$ for $BaFe_2(S_{0.75}Se_{0.25})_3$. The $T_{co}$ value is lower than both values of $BaFe_2S_3$ and $BaFe_2Se_3$ and the superconductivity is observed in the limited pressure range 10–12 GPa. The $T_{co}$ value is not enhanced in $BaFe_2(S_{0.75}Se_{0.25})_3$ despite the suppression of the $T_N$ and the $T^*$, and the decrease of resistance at superconducting transition is not clearly observed. These are likely because some amount of magnetic ordering remains under high pressure. The stripe-type magnetic ordering in $BaFe_2(S_{0.75}Se_{0.25})_3$ seems to remain under high pressure because of the structural phase transition to the orthorhombic $Cmcm$ similar to $BaFe_2Se_3$ which exhibits structural transition from $Pmn2_1$ to $Cmcm$ at 4 GPa at ambient temperature [29].

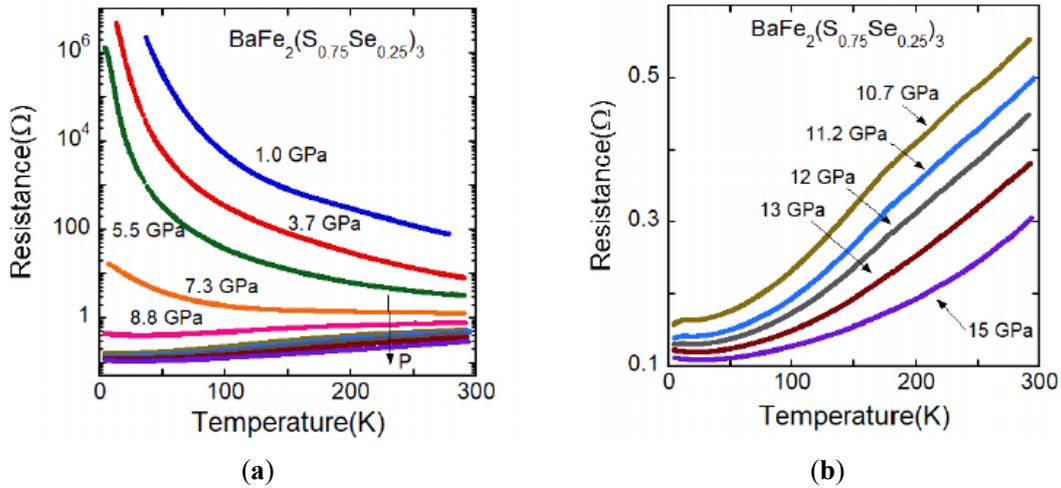

(a)          (b)

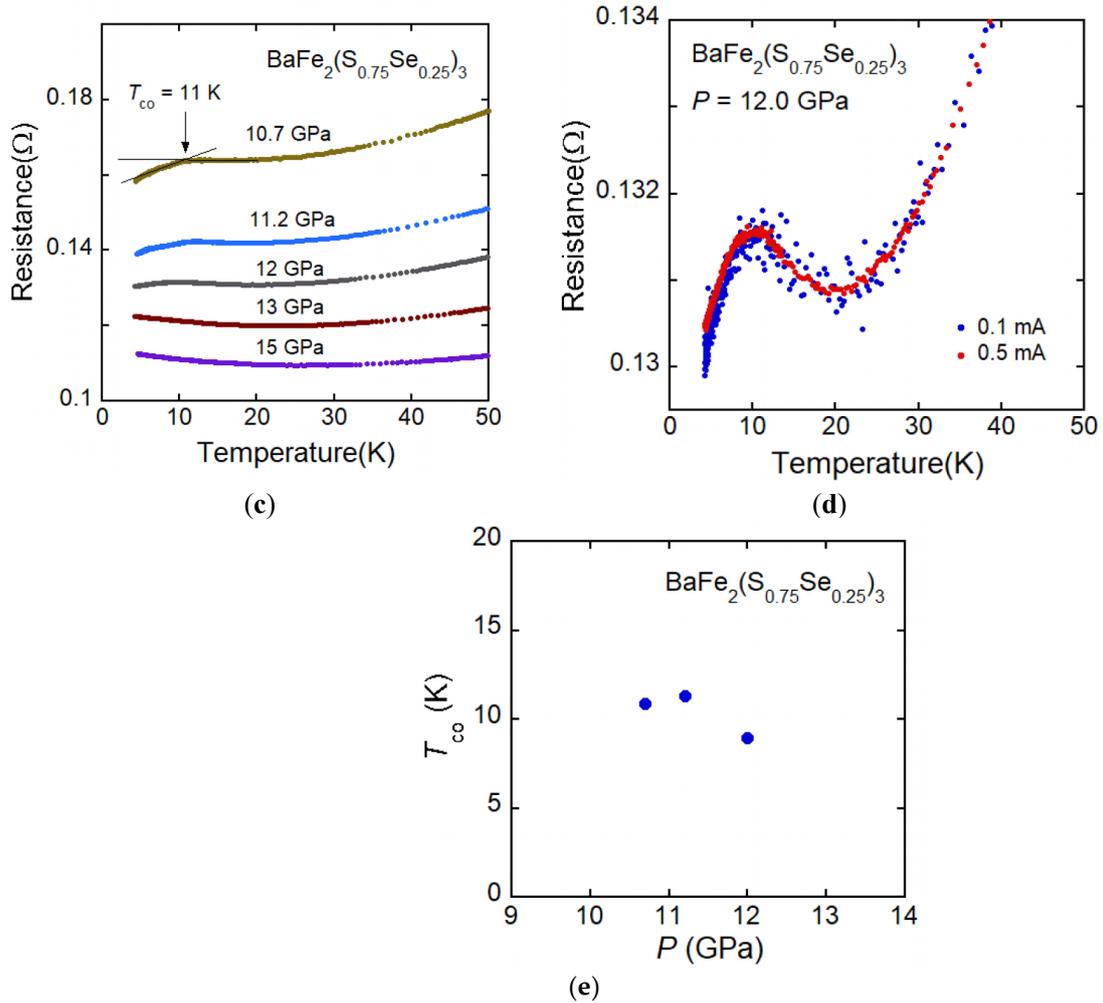

Figure 5. (a) Temperature dependence of the electrical resistance of BaFe$_2$(S$_{0.75}$Se$_{0.25}$)$_3$ along the leg direction. Insulating behavior is gradually suppressed and the metal-insulator transition is observed around 7–10 GPa. (b) $R(T)$ curve of metallic state for BaFe$_2$(S$_{0.75}$Se$_{0.25}$)$_3$. (c) $R(T)$ curve in the low-temperature range. A decrease of resistance around 10 K is observed, which is attributed to the superconductivity. (d) The superconducting transition at 12 GPa shifts to the low-temperature side with increasing electric current, which is consistent with superconducting characteristics. (e) Pressure dependence of $T_{co}$ for BaFe$_2$(S$_{0.75}$Se$_{0.25}$)$_3$. The superconductivity is observed in the limited pressure range 10–12 GPa.

*3.2. Electrical Resistivity under High Pressure for Spin-Ladder Compound BaFe$_2$Se$_3$*

We performed electrical resistance measurements under high pressure for three BaFe$_2$Se$_3$ samples. Figure 6a shows the $R(T)$ curve of BaFe$_2$Se$_3$ (sample 2) along the leg direction. Insulating behavior observed at ambient pressure caused by the electron correlation effect is gradually suppressed and the $R(T)$ curve exhibits a metal-insulator transition at ~9 GPa. Figure 6b shows the $R(T)$ curve of the metallic state and one can see a sudden decrease in resistance around 13 K at 11 GPa. This resistance decrease is thought to be attributable to the superconducting transition. As shown in the inset of Figure 6b, the superconducting transition at 12.2 GPa shifts to the low-temperature side with increasing electric current, which is consistent with superconducting characteristics. According to the preceding study [22], the superconductivity with $T_{co}$ = 11 K was observed at 11.5 GPa, which is almost the same result as our experiment.

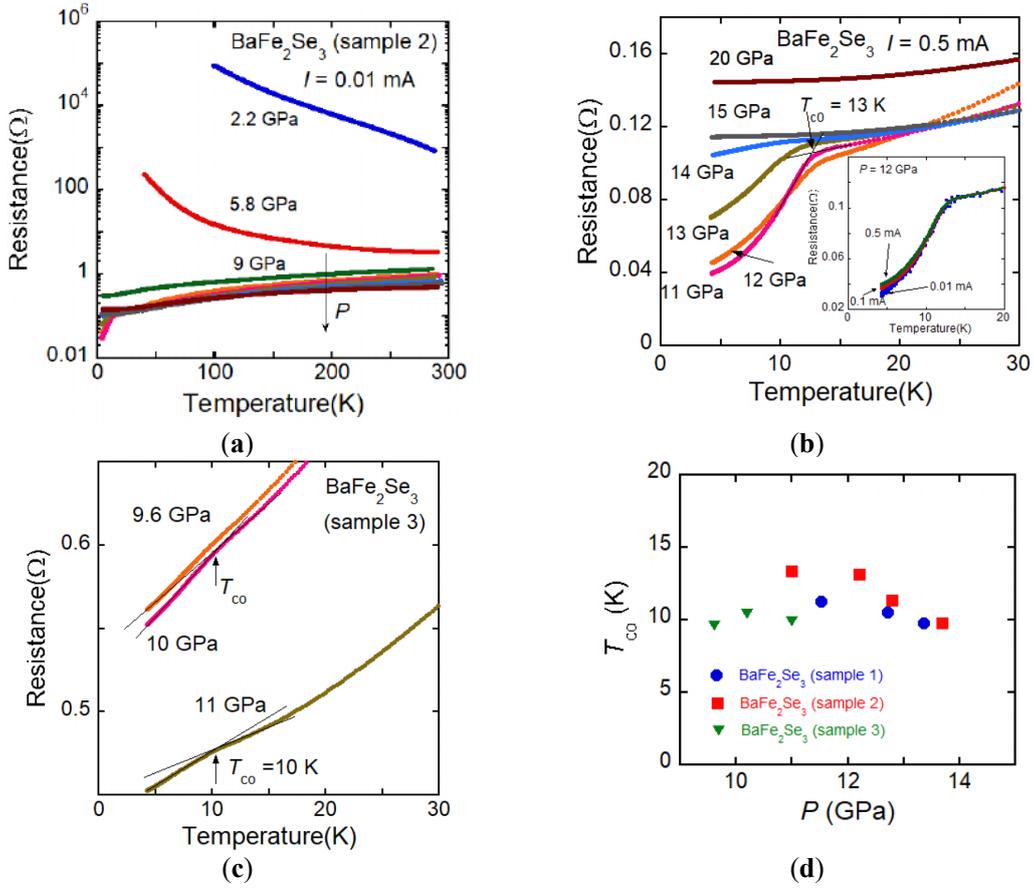

**Figure 6.** (**a**) Temperature dependence of the electrical resistance of BaFe₂Se₃ along the leg direction. Insulating behavior is gradually suppressed and the metal-insulator transition is observed around 9 GPa. (**b**) $R(T)$ curve of metallic state for BaFe₂Se₃. The superconductivity is observed at 11 GPa and suppressed with further compression. The inset shows the current dependence of $R(T)$ curve. The superconducting transition at 12 GPa shifts to the low-temperature side with increasing current, which is consistent with superconducting characteristics. (**c**) $R(T)$ curve of metallic state for BaFe₂Se₃ under a more hydrostatic condition using DAC. A slight decrease of resistance around 10 K is observed, which is attributed to the superconductivity. (**d**) The pressure dependence of $T_{co}$ is exhibited. It seems that a hydrostatic condition is not always advantageous to the appearance of superconductivity.

The high-pressure measurements for samples 1 and 2 were carried out using NaCl as a pressure transmitting medium (PTM) and the results of sample 1 are almost the same as the results of sample 2. Since the sample space of the DAC is very tiny, it is difficult to make electric contact using conductive paste between sample and lead. Then the sample embedded in the soft material (NaCl) and the electric contact between lead and sample is directly pressed by the anvil. Due to the loss of the hydrostatic condition by use of solid PTM, zero resistance could not be observed. In order to improve the hydrostatic condition, high-pressure measurements were performed by making a NaCl layer between the sample and the anvil and then the sample is completely covered with NaCl. Figure 6c shows the $R(T)$ curve taken under such a condition for BaFe₂Se₃ (sample 3). As shown in Figure 6c, a metallic behavior was observed above 9.6 GPa, and a slight decrease of resistance was observed around 10 K. The superconducting transition at 11 GPa shifts to the low-temperature side with increasing electric current, which is consistent with superconducting characteristics. Therefore, this slight decrease of resistance is attributed to superconductivity. These results indicate that anisotropic pressure is advantageous to the appearance of superconductivity in BaFe$_2$Se$_3$. The pressure dependence of $T_{co}$ is summarized in Figure 6d.

*3.3. Electrical Resistivity under High Pressure for Spin-Ladder Compound BaFe$_{2+\delta}$Se$_3$*

Figure 7a shows the temperature dependence of the electrical resistance of BaFe$_{2.1}$Se$_3$ along the leg direction above the pressure of metal-insulator transition around 10 GPa and Figure 7b shows the $R(T)$ curve at low temperature. The decrease of resistance caused by superconductivity is observed around 10 K. Figure 8a shows the $R(T)$ curve of BaFe$_{2.2}$Se$_3$ along the leg direction above the pressure of metal-insulator transition around 13 GPa and Figure 8b shows the $R(T)$ curve at low temperature. Figure 8c shows the pressure dependence of $T_{co}$ for BaFe$_2$Se$_3$, BaFe$_{2.1}$Se$_3$, and BaFe$_{2.2}$Se$_3$. For BaFe$_{2.1}$Se$_3$, the superconductivity is observed in a higher-pressure range, 13–17 GPa than the ones of BaFe$_2$Se$_3$ and BaFe$_{2.1}$Se$_3$. The $T_N$ values are 242 K for BaFe$_2$Se$_3$, 241 K for BaFe$_{2.1}$Se$_3$, and 245 K for BaFe$_{2.2}$Se$_3$. Since the $T_N$ value of 245 K for BaFe$_{2.2}$Se$_3$ is higher than the ones of BaFe$_2$Se$_3$ and BaFe$_{2.1}$Se$_3$, it seems that higher pressure is necessary for BaFe$_{2.2}$Se$_3$ to suppress the magnetic ordering and induce superconductivity.

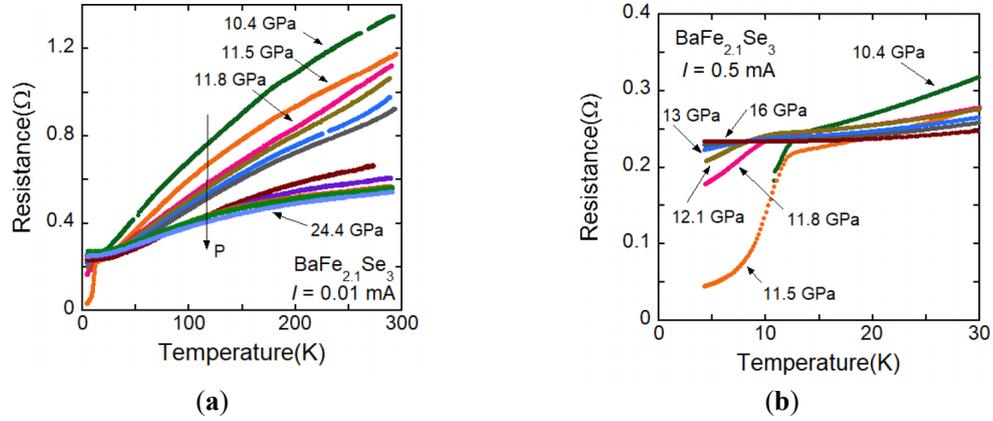

**Figure 7.** (**a**) $R(T)$ curve of BaFe$_{2.1}$Se$_3$ along the leg direction above the pressure exhibiting metal-insulator transition around 10 GPa (**b**) $R(T)$ curve at low temperature. The decrease of resistance caused by superconductivity is observed at 10–13 K.

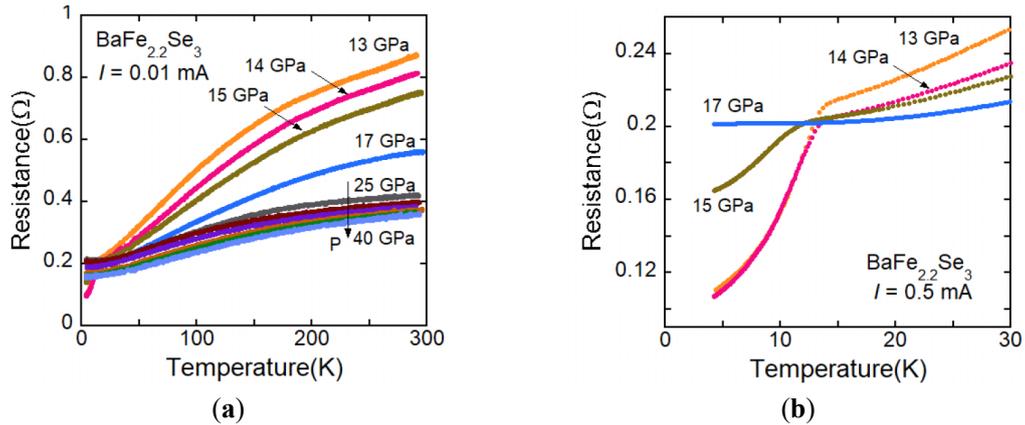

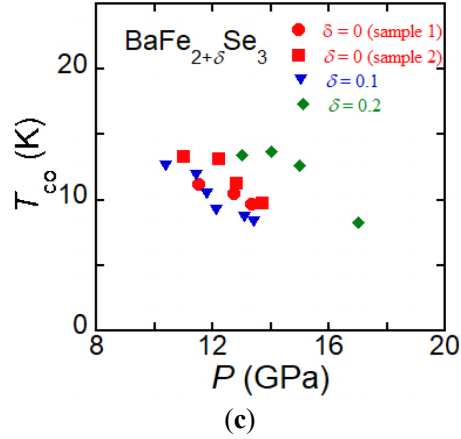

(c)

**Figure 8.** (**a**) $R(T)$ curve of BaFe$_{2.2}$Se$_3$ along the leg direction above the pressure exhibiting metal-insulator transition around 13 GPa. (**b**) $R(T)$ curve at low temperature. The decrease of resistance caused by superconductivity is observed at 14 K. (**c**) Pressure dependence of $T_{co}$ for BaFe$_{2.1}$Se$_3$, BaFe$_{2.1}$Se$_3$, and BaFe$_{2.2}$Se$_3$. Higher pressure is necessary for BaFe$_{2.2}$Se$_3$ to induce superconductivity.

It is indicated that the pressure-induced superconductivity of BaFe$_2$Se$_3$ appears with existing magnetic moment, which is different from the case of BaFe$_2$S$_3$. Theoretical calculation suggests that superconductivity of BaFe$_2$S$_3$ occurs when magnetic moment vanishes, while in BaFe$_2$Se$_3$, superconductivity appears with magnetism. Since it is experimentally and theoretically shown all magnetic moment vanishes at 30 GPa [22,30], high-pressure resistance measurements were performed to 30 GPa. The $R(T)$ curves in the high-pressure range for BaFe$_{2.1}$Se$_3$ and BaFe$_{2.2}$Se$_3$ are shown in Figure 9a,b, respectively. However, no anomalous behavior related to the superconductivity was observed at low temperature. Since the magnetic ordering could not be detected in the high-pressure measurement, it could not be confirmed that the magnetism disappears at 30 GPa at low temperature. It is otherwise likely that no superconductivity might appear without magnetic moment. It is needed to characterize the magnetism above 30 GPa at low temperature.

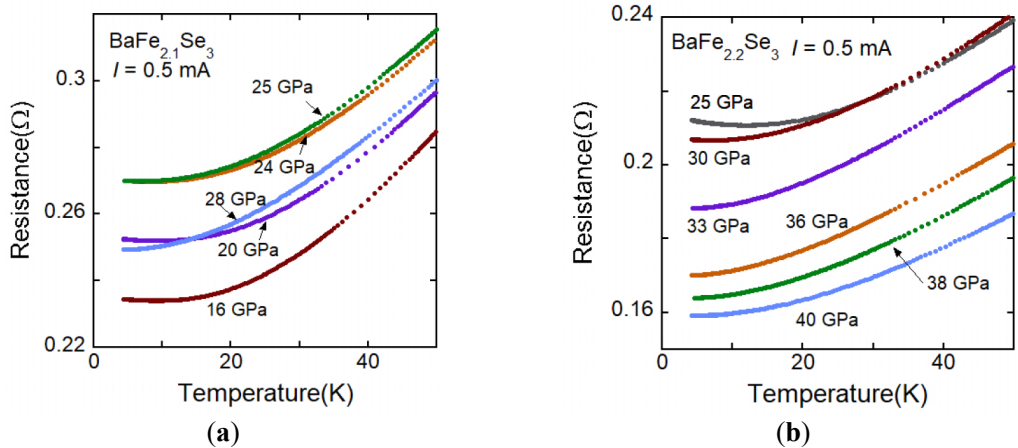

**Figure 9.** (**a**) $R(T)$ curve for BaFe$_{2.1}$Se$_3$ under high pressure to about 30 GPa, where superconductivity is predicted by the theoretical calculation [30]. No anomalous behavior related to the superconductivity was observed. (**b**) $R(T)$ curve for BaFe$_{2.2}$Se$_3$ under high pressure to 40 GPa. No anomalous behavior related to the superconductivity was observed, either.

## 4. Conclusions

In this study, we have performed electrical resistivity measurements for single crystalline iron-based spin-ladder compounds BaFe$_2$(S$_{0.75}$Se$_{0.25}$)$_3$ and BaFe$_{2+\delta}$Se$_3$ (**$\delta$ = 0, 0.1, and 0.2**). We studied the superconductivity for the composition in the vicinity of the phase boundary between stripe-type and block-type magnetic ordering in BaFe$_2$(S$_{1-x}$Se$_x$)$_3$. The superconductivity can be detected as a small decrease of resistance at ~10 K in the pressure range of 10–12 GPa. However, the $T_c$ of ~ 10 K is slightly smaller than BaFe$_2$S$_3$ and BaFe$_2$Se$_3$, and the superconductivity appears in the limited pressure range of 10–12 GPa. This is likely because some amount of stripe-type magnetic ordering remains under the pressure where the superconductivity is observed. We confirmed the superconductivity of BaFe$_2$Se$_3$ at 11 K and 11.5 GPa as with the preceding report [22]. We also observed anisotropic pressure is advantageous to the appearance of superconductivity i**n** BaFe$_2$Se$_3$. **S**ince the $T_{co}$ changes with similar trend to the value of $T_N$ in BaFe$_2$S$_3$, we examined the effect of the stoichiometry on the pressure-induced superconductivity f**or** BaFe$_{2+\delta}$Se$_3$ ($\delta$ = 0, 0.1, and 0.2)**.** For BaFe$_{2+\delta}$Se$_3$ ($\delta$ = 0, 0.1, and 0.2)**, th**ough the $T_N$ value for BaFe$_{2.2}$Se$_3$ was higher than the ones of BaFe$_2$Se$_3$ and BaFe$_{2.1}$Se$_3$, a higher value of $T_{co}$ was not observed in BaFe$_{2.2}$Se$_3$. The pressure range where the superconductivity appeared shifted to a higher-pressure range for BaFe$_{2.2}$Se$_3$. These results indicate that enough pressure to suppress the effect of magnetism is necessary for BaFe$_{2.2}$Se$_3$ to exhibit superconductivity. Finally, superconductivity was not observed around 30 GPa, at which the magnetic moment is thought to disappear. Further high-pressure studies for these spin-ladder systems are needed to understand the interplay between magnetism, orbital ordering, and superconductivity.

**Author Contributions:** Conceptualization, H.T., T.A. and K.O.; Formal analysis, H.T. and R.K.; Funding acquisition, H.T., T.A. and K.O.; Investigation, H.T., R.K., Chizuru Kawashima and T.A.; Project administration, H.T.; Resources, S.I., T.A. and K.O.; Supervision, H.T.; Visualization, H.T. and C.K.; Writing–original draft, H.T.; Writing–review & editing, K.O. All authors have read and agreed to the published version of the manuscript.

**Funding:** This research was funded by JSPS, Grants-in-Aid for Scientific Research JP20K14396, JP19H05823, JP19H05822, JP18H01159, and JP16H04019. This work was also supported by Nihon University Multidisciplinary Research Grant for 2016–2017, Nihon University President Grant Initiative (2018–2020), and JST CREST Grant No. JP19198318, Japan.

**Conflicts of Interest:** The authors declare no conflict of interest.